\documentclass[twocolumn,           
               showpacs,            
               preprintnumbers,     
               aps,                 
               prl,                 
               a4paper,             
               superscriptaddress,  
               nofootinbib,         
               tightenlines,        
               floats,floatfix      
               ]{revtex4}

\usepackage{graphicx}
\usepackage{subfigure}
\usepackage{latexsym}
\usepackage{amsmath,amssymb}        
\usepackage[draft=false]{hyperref}

\input colordvi

\begin{document}



\title{Inflation, dark matter and dark energy in the string landscape}
\author{Andrew R.~Liddle}
\affiliation{Astronomy Centre, University of Sussex, Brighton BN1 9QH,
United Kingdom}
\author{L. Arturo Ure\~na-L\'opez}
\affiliation{Instituto de F\'isica de la Universidad de Guanajuato,
C.P.~37150, Le\'on, Guanajuato, M\'exico}  
\date{\today}
\pacs{98.80.-k, 98.80.Cq \hfill astro-ph/0605205}
\preprint{astro-ph/0605205}


\begin{abstract}
We consider the conditions needed to unify the description of dark
matter, dark energy and inflation in the context of the string
landscape. We find that incomplete decay of the inflaton field gives
the possibility that a single field is responsible for all three
phenomena. By contrast, unifying dark matter and dark energy into a
single field, separate from the inflaton, appears rather difficult.
\end{abstract}

\maketitle


\section{Introduction}

It has become clear recently that potentials of the form $V_0 +
\frac{1}{2} m^2 \phi^2$, where $V_0$ has the small value needed to
explain the observed dark energy density, are plausibly motivated by a
combination of the string landscape picture and the anthropic
principle, and are not necessarily hopelessly fine-tuned as previously
thought. The gist of the argument \cite{sland} is that string
theory contains a huge number of possible configurations of differing
vacuum energy, which might be exhaustively explored throughout the
very large scale Universe, for instance via a self-reproducing
inflationary cosmology mechanism \cite{self}. The selection
effect that we must live in a region of the Universe capable of
forming stars and galaxies then enforces that we live in a region
where $V_0$ is atypically small, but non-zero. With some caveats
\cite{TARW}, this picture gives an impressive probabilistic prediction
of the order of magnitude of the dark energy density
\cite{anth}.

Potentials of the form above are of interest as they offer the
possibility of a unified description of various features of the
Universe for which scalar fields have been invoked, specifically
inflation, dark matter, and dark energy. The main ingredients to do
this are already in the literature, though they have not been
explicitly connected. In their work on post-inflationary preheating,
Kofman et al.~\cite{Kofman:1994rk} remarked that the inflaton decay
might be incomplete, with the residue having the capability of acting
as dark matter. Separately, Linde has noted \cite{Linde} that with $m
\simeq 10^{-6} m_{{\rm Pl}}$, the above potentials unify standard
chaotic inflation (during which $V_0$ is utterly negligible) with dark
energy. The precise form of the non-constant part of the potential is
not of course crucial to this argument; any of the normal inflationary
potentials will achieve the same once $V_0$ is added.\footnote{It is
not really accurate to associate $V_0$ with a particular field: the
vacuum energy is a property of the full Lagrangian. Nevertheless, in
the landscape picture it is useful to think of it in these terms.}

In this paper we wish to consider the possible additional unification
of cold dark matter (CDM) with dark energy using such
potentials. While usual particle dark matter candidates such as
weakly-interacting massive particles (WIMPs) correspond to incoherent
distributions of individual particles, it has long been known that an
alternative CDM candidate is a coherently oscillating scalar field,
the archetypal example being axion dark matter. Provided the potential
is of quadratic form about its minimum, such a field behaves on
average like pressureless matter \cite{Kolb:1990vq}, and is
indistinguishable from traditional CDM candidates provided the
oscillation period is much shorter than any other dynamical scale in
the problem (true unless the field is super-light). Furthermore, it is
known that linear perturbations in such a coherent field mimic those
of a pressureless fluid \cite{Hwang:1996xd}, and that non-linear
top-hat collapse proceeds in the same way. Such coherent scalar fields
are therefore a well developed alternative to the WIMP paradigm.

We do not aim to make any specific proposals for how such unified
scenarios might arise from fundamental theories, but rather wish to
explore what conditions would have to be met in order for such
scenarios to be compatible with observations. We explore two types of
scenario:
\begin{enumerate}
\setlength{\itemsep}{0pt}
\item Unification of inflation, dark matter and dark energy into the
same scalar field $\phi$.
\item Unification of dark matter and dark energy into a single scalar
field $\phi$, with inflation provided by a separate scalar field
$\psi$.
\end{enumerate}
The main conditions that will concern us is whether a complete history
of the Universe from inflation onwards can be constructed, with the
fields taking plausible values, and whether perturbations can be
generated that are compatible with the observation that isocurvature
perturbations, if present at all, are subdominant to adiabatic ones.

There have been many attempts to use scalar fields to unify
combinations of inflation, dark matter, and dark energy.
For instance, Ref.~\cite{Padmanabhan:2002sh} proposed a tachyon-type
scalar-field Lagrangian, in which the scalar fluid can be
broken up into dark matter and dark energy components. A k-essence
unification of dark matter and dark energy was given in
Ref.~\cite{Scherrer:2004au}. Staying instead with the canonical
Lagrangian, Ref.~\cite{Arbey:2006it} introduced a complex scalar field
with a mixed potential made of quadratic and exponential terms, which
then mimic dark matter and dark energy, respectively, at the scales of
interest. Unification scenarios featuring inflation include
quintessential inflation (unifying inflation+dark energy)
\cite{Peebles:1998qn}, inflaton+dark matter in the braneworld scenario
\cite{Lidsey:2001nj}, and braneworld inflaton+dark matter+cosmological
constant from multiple fields in a type IIB supergravity theory
\cite{Matos:2005yt}.  Our proposal is a simpler one than any of those
listed above, with Ref.~\cite{Matos:2005yt} being the closest.

Scalar fields in quadratic potentials have a generic evolution.
Initially, while $m \ll H$ (where $H$ is the Hubble parameter), the
scalar field is frozen by the friction of the expanding Universe and
remains constant, corresponding to a constant energy density. If at
that time the field is the dominant energy density in the Universe it
will drive inflation. Once $H$ falls below $m$ the scalar field begins
to oscillate, and its time-averaged evolution has density $\rho_\phi$
falling as $1/a^3$, exactly as CDM \cite{Kolb:1990vq}. The
normalization of the density is determined by the initial amplitude of
the scalar field oscillations, as follows.

In order to recover the standard dark matter scenario, the scalar mass
should satisfy $m \gg H_{{\rm eq}}$, where $H_{{\rm eq}}$ is the
Hubble parameter at the time of radiation and matter equality, so the
oscillations of the field begin well within the radiation-dominated
era. If we denote by $t_\ast$ the time at which the scalar mass equals
the Hubble parameter, $m=H_\ast$, then $m^2 = 8 \pi \rho_{\rm R*}/3
m^2_{{\rm Pl}}$, where $\rho_{\rm R}$ is energy density of
relativistic matter and $m_{{\rm Pl}}$ is the Planck mass. The photon
density is related to the total radiation density by $\rho_{{\rm R}} =
(g/2) \rho_\gamma$ where $g$ is the total number of relativistic
particle degrees of freedom \cite{Kolb:1990vq}.

The averaged scalar field energy density is given by $\rho_\phi =
\frac{1}{2} m^2 \phi^2_\ast a_\ast^3 /a^3$ for $t > t_\ast$; here
$\phi_\ast$ is the initial scalar field amplitude at $t_\ast$. We
define the scalar field dark matter mass per photon as $\xi_{{\rm dm}}
\equiv \rho_\phi/n_\gamma$.  This quantity is constant for $t_\ast <
t$ apart from changes in the number of relativistic species; we assume
expansion at constant entropy implying that $\xi/g_{\rm S}$ remains
constant where $g_{\rm S}$ is the entropic degrees of freedom, usually
very similar to $g$ \cite{Kolb:1990vq}.  Using `0' to indicate present
values, the present scalar field dark matter mass per photon is then
\begin{equation}
\xi_{{\rm dm},0} \simeq 4 \, \frac{g_{{\rm S},0}}{g_*^{1/4}}
    \left( \frac{m}{m_{{\rm Pl}}} \right)^{1/2}
    \frac{\phi_\ast^2}{m_{{\rm Pl}}^2} \; m_{{\rm Pl}}  \,.
\label{eq:1}
\end{equation}

The measured value of the current dark matter mass per photon is
$\xi_{{\rm dm},0}=2.2 \times 10^{-28} m_{{\rm Pl}}$ using values from
WMAP3 \cite{Spergel:2006hy}, which for typical values $g_* \simeq
100$, \mbox{$g_{{\rm S},0} = 3.9$} then gives the following constraint
\begin{equation}
\left(\frac{m}{m_{{\rm Pl}}} \right)^{1/2} \frac{\phi_\ast^2}{m_{{\rm
    Pl}}^2} \simeq  4 \times 10^{-29}\, . \label{eq:2}
\end{equation}

A lower limit can be placed on $m$ from structure formation.  The
linearly-perturbed scalar field equation resembles a damped and forced
harmonic oscillator. For scales with a comoving wavenumber $k < a(t)
m$, where $a(t)$ is the scale factor, the field perturbation is in
resonance with the force term. In this case, the field's density
contrast grows as that of CDM \cite{Hwang:1996xd}.  However, for $k >
a(t) m$ the perturbed field is out of phase with the force term, and
then the perturbations are suppressed relative to the standard CDM
case. The \emph{largest} scale at which suppression occurs corresponds
to the \emph{smallest} scale factor; in our case that scale is $k_\ast
= a_\ast m$. Assuming the same conditions that led to
Eq.~(\ref{eq:1}), together with the restriction $k_\ast > 1 \, {\rm
Mpc}^{-1}$, we get the lower bound $m/m_{\rm Pl} > 7 \times 10^{-52}$,
i.e.~$m > 10^{-23} \, {\rm eV}$. Provided the field is significantly
more massive than this, it will behave indistinguishably from standard
CDM. If instead it more or less saturates this bound, the Compton
wavelength of the particles may become comparable to astrophysical
scales with observable consequences (see
Refs.~\cite{Sahni:1999qe,Hu:2000ke} and references therein).

\section{Triple unification: inflation, dark matter and dark energy
from a single field}

In this section, we explore the conditions needed to unify all three
phenomena --- dark matter, dark energy, and inflation --- into a
single field. For simplicity we will assume that the quadratic form of
the potential holds for all relevant $\phi$ values, though other
choices can be made.

The advantage of the single-field unified scenario is that the only
perturbations generated during inflation are adiabatic, as that is the
only type that a single field can support. Obtaining the correct
amplitude of scalar primordial perturbations requires $m/m_{\rm Pl}
\simeq 10^{-6}$, and the spectral index is independent of $m$ and a
good fit to WMAP3 data \cite{Spergel:2006hy}.  However at the end of
inflation $\phi$ is still of order of $m_{{\rm Pl}}$. By contrast,
Eq.~(\ref{eq:2}) requires an initial amplitude for the dark
matter oscillations of the order of $\phi_\ast \simeq 10^{-13} m_{{\rm
Pl}}$. The main requirement for a working scenario therefore is a
drastic but incomplete reduction of the amplitude of the inflaton
oscillations during reheating, reducing the energy density of the
inflaton field by a factor of about $10^{26}$. This is necessary to
permit a long radiation-dominated epoch.

Such an incomplete decay indicates that the reheating mechanism
should be via inflaton annihilations, rather than decays. This is,
for instance, guaranteed to be true if the reflection symmetry of
the inflaton potential is not spontaneously broken, as then only
quadratic interaction terms are permitted.  In such circumstances it
is generically true that there will be some residual inflaton
density left over, because once the density becomes low enough the
particles are no longer able to `find' each other to annihilate
\cite{Kofman:1994rk}.  The question is whether a mechanism can be
found which reduces the inflaton density by the amount required by
the considerations above.

The two main paradigms for conversion of the inflaton into other
matter are preheating (coherent multi-particle decays) and reheating
(single particle decays), which may happen in sequence.  Some
mechanisms for the decay of the inflaton have been proposed in the
literature, see for instance Refs.~\cite{Abbott:1986kb,Kolb:1990vq,
Kofman:1994rk,Shtanov:1994ce,Kofman:1997yn,Podolsky:2005bw} and
references therein. The relevance of (p)reheating to unification
scenarios has been discussed in Ref.~\cite{uniheat}.

The conventional reheating mechanism, corresponding to single-particle
decays, adds a constant decay width $\Gamma_\phi$ to the inflaton
equation of motion \cite{Abbott:1986kb,Kolb:1990vq} in the form
\begin{equation}
  \dot{\rho}_\phi + 3H \dot{\phi}^2 = -\Gamma_\phi \dot{\phi}^2
  \,. \label{eq:4}
\end{equation}
Eq.~(\ref{eq:4}) implies the usual exponential decay law for scalar
particles which are linearly coupled to other bosons and fermions
\cite{Linde:2005ht,Kolb:1990vq}. It proceeds once $\Gamma_\phi \gg H$,
and leads to complete decay of the scalar field. Conventional
reheating can play an important role in reducing the inflaton energy
density as required by the triple unification scenario, but needs
modification to prevent the decay being complete.

By contrast, preheating offers a mechanism for rapid but incomplete
decay of the inflaton field, provided the inflaton is coupled to
another scalar field $\chi$ through simple four-legs interactions of
the form $g^2 \phi^2 \chi^2$, where $g$ is the coupling constant. That
this could make the inflaton field a dark matter candidate was first
noted by Kofman et al.~\cite{Kofman:1994rk}, though they did not
evaluate in detail the conditions needed to realize this. Further
analysis of the scenario can be found in
Refs.~\cite{Shtanov:1994ce,Kofman:1997yn,Greene:1997fu,
Podolsky:2005bw}.

The conclusion of that work is that the decay is indeed incomplete,
with preheating coming to an end once the amplitude of the inflaton
oscillations becomes smaller than $m/g$
\cite{Kofman:1994rk,Shtanov:1994ce}.  While this does give a large
reduction provided $g$ is not too small, the amplitude required for
CDM, Eq.~(\ref{eq:2}), is $\phi_\ast \sim 10^{-7} m$, and hence we
would need $g \sim 10^7$. Such a non-perturbatively large coupling is
unattractive, even if supersymmetry is invoked to cancel radiative
corrections \cite{Podolsky:2005bw}. A further problem
\cite{Kofman:1997yn} is that the density of $\chi$-particles produced
may be less than that of incoherent inflaton particles, which prevents
generation of a satisfactory radiation-dominated era.

The efficiency of preheating is enhanced if one includes a linear
(three-leg) coupling between the inflaton field and other bosonic and
fermionic fields \cite{Kofman:1997yn,Shtanov:1994ce}. This however
makes the inflaton field decay completely \cite{Podolsky:2005bw},
contrary to our aim.

In conclusion, preheating sets the precedent of incomplete inflaton
decay, but existing models do not satisfy the conditions needed by the
triple unification scenario; quadratic interactions give too little
decay and linear ones too much. It may therefore be necessary to
exploit annihilations via perturbative interactions. As a simple toy
model, consider Eq.~(\ref{eq:4}) but with the decay width now allowed
to depend on the scalar field density; for instance $\Gamma_\phi
\propto \rho_\phi$ corresponds to two-body annihilations (`decay' rate
proportional to the local density). This alone is insufficient as the
annihilations would be important during inflation; a viable form would
be
\begin{equation}
\Gamma_\phi = \Gamma_0 \frac{\rho_\phi/\rho_{{\rm
c}}}{1+\rho_\phi/\rho_{{\rm c}}} \,,
\end{equation}
which makes a smooth transition from single particle decays to two-body
annihilations as $\rho_\phi$ reduces. With suitable tuning of the constants
$\Gamma_0$ and $\rho_{{\rm c}}$ a viable scenario can be constructed,
though this form of the decay width is not motivated by any fundamental
considerations. 

To end this section, we note that it is by no means essential for the
potential to take the form $V_0 + \frac{1}{2} m^2 \phi^2$ all the way
up to the $\phi$ values responsible for inflation; see for instance
Refs.~\cite{Peebles:1999fz,Sahni:1999qe}. The unification of dark
matter and dark energy only requires that it is of this form for very
small $\phi$. Indeed, in the context of the string landscape, and
bearing in mind the spectral index measurements from WMAP3
\cite{Spergel:2006hy}, it may be more natural that inflation takes
place near a maximum of the potential \cite{Alabidi:2006qa}, perhaps
with initial conditions fixed by the topological inflation mechanism
\cite{topinf}.  Clearly it would be interesting to explore incomplete
decay mechanisms for a range of inflationary potentials.

\section{Unification of dark matter and dark energy into a single
field}

In this section we consider what appears to be a less ambitious
scenario, where only dark matter and dark energy are unified by the
$\phi$ field, with some other field responsible for inflation. This
has some similarities to the curvaton scenario, but is more
restrictive since the $\phi$ field {\em is} the dark matter, rather
than decaying into the dark matter. The mass of the field is now not
directly determined by the perturbation normalization, and instead
should have a sufficiently small value to avoid interfering with the
inflaton, $m^2 \phi_*^2 \ll m_{{\rm Pl}}^2 H^2$. In fact, this
scenario proves rather hard to achieve.

The first problem encountered by such scenarios would be to explain
the small value of $\phi_*$ required by Eq.~(\ref{eq:2}), since there
appears no reason why the field should be so close to its
minimum. Furthermore, this initial condition must not be spoiled by
quantum fluctuations induced in $\phi$ during inflation, which are of
order $H/2\pi$ per $e$-folding. Indeed these fluctuations must be
small enough that the primordial CDM perturbation does not exceed the
observed $10^{-5}$ value. This imposes the tight condition
\begin{equation}
\frac{\delta \phi}{\phi_*} \simeq \frac{H}{2\pi \phi_*} \lesssim
10^{-5} \,.
\end{equation}

Inflation generates the correct amplitude of perturbations provided
$H/m_{{\rm Pl}} \simeq 10^{-4} \epsilon^{1/2}$, where $\epsilon < 1$
is the slow-roll parameter for the inflaton. Since the observable
perturbations may come from $\phi$ rather than the inflaton, this is
an upper limit on $H$. The combination of these constraints with
Eq.~(\ref{eq:2}) gives the powerful limit $m \lesssim \epsilon^{-2}
\times 5 \times 10^{-30} \, {\rm eV}$.  For a viable scenario
satisfying the lower mass limit from structure formation quoted
earlier, this forces $\epsilon$, and hence the inflationary energy
scale, to be very low, and even then the scalar mass is forced to be
extremely light.  Additionally, the appropriate initial value of
$\phi_*$ must arise by accident (this is also a feature of the
curvaton scenario).

Even if these circumstances are satisfied, there is a further problem
that perturbations in the CDM arise from separate fluctuations to
those in the baryon--photon fluid. They are therefore of isocurvature
form, which is highly disfavoured by data if the adiabatic
perturbations are negligible. This issue has typically been ignored in
previous attempts to unify dark matter and dark energy. Perhaps the
scenario can be saved by allowing the inflaton perturbations to be
non-negligible, thus giving a mixture of adiabatic and (partially
correlated) isocurvature perturbations, but previous studies are not
encouraging \cite{Polarski:1994rz}. Another possible escape would be
if the $\phi$ field is at least partly excited by the inflaton decay,
giving it an adiabatic perturbation (c.f.~the mention of trace
decoupled CDM in curvaton decays in Ref.~\cite{Lyth:2001nq}). Note
this must be a coherent excitation generating a universal mean value
$\phi_0$ about which small perturbations are superimposed via the
adiabatic perturbations, requiring a breaking of the potential's
reflection symmetry in the interactions between the inflaton and
$\phi$.

\section{Conclusions}

We have examined the possibility that ideas coming from the string
landscape can unify various key aspects of cosmology into a single
field, specifically inflation, dark matter and dark energy. We have
not been able to be very specific in terms of particle physics models,
but we have investigated the general conditions necessary to bring
about such a unification. Curiously, scenarios unifying all three
phenomena appear to be easier to realize than those which keep a
separate inflaton. The key ingredient required to make such scenarios
a reality is partial, nearly complete, decay of the inflaton into the
baryon--radiation fluid, so that the residual decoupled component can
survive as dark matter and dark energy. Preheating scenarios may offer
such a possibility \cite{Kofman:1994rk}.

\vskip 10pt

\begin{acknowledgments}
A.R.L.\ was supported by PPARC and L.A.U.-L.\ by CONACYT (42748,
46195, 47641), CONCYTEG 05-16-K117-032, DINPO and PROMEP-UGTO-CA-3. A
visit to Sussex by L.A.U.-L., funded by the Royal Society and the
Academia Mexicana de Ciencias, initiated this work. We thank Ed
Copeland, James Lidsey, Tonatiuh Matos, C\'edric Pahud, and David
Wands for discussions.
\end{acknowledgments}



\end{document}